# The Human Plasma Membrane Peripherome: Visualization and Analysis of Interactions

Katerina C. Nastou, Georgios N. Tsaousis, Kimon E. Kremizas, Zoi I. Litou and Stavros J. Hamodrakas[*]

Department of Cell Biology and Biophysics, Faculty of Biology, University of Athens, Panepistimiopolis, Athens 15701, Greece

*To whom correspondence should be addressed

Prof. Stavros J. Hamodrakas

Department of Cell Biology and Biophysics, Faculty of Biology,

University of Athens, Panepistimiopolis,

Athens 15701, Greece

Phone: +30 210 727 4931

Fax: +30 210 7274254

E.mail: shamodr@biol.uoa.gr

http://biophysics.biol.uoa.gr

**ABSTRACT**

A major part of membrane function is conducted by proteins, both integral and peripheral. Peripheral membrane proteins temporarily adhere to biological membranes, either to the lipid bilayer or to integral membrane proteins with non-covalent interactions. The aim of this study was to construct and analyze the interactions of the human plasma membrane peripheral proteins (peripherome hereinafter). For this purpose, we collected a dataset of peripheral proteins of the human plasma membrane. We also collected a dataset of experimentally verified interactions for these proteins. The interaction network created from this dataset has been visualized using Cytoscape. We grouped the proteins based on their subcellular location and clustered them using the MCL algorithm in order to detect functional modules. Moreover, functional and graph theory based analyses have been performed to assess biological features of the network. Interaction data with drug molecules show that ~10% of peripheral membrane proteins are targets for approved drugs, suggesting their potential implications in disease. In conclusion, we reveal novel features and properties regarding the protein-protein interaction network created by peripheral proteins of the human plasma membrane.

# 1. INTRODUCTION

Since the completion of the human genome project in 2001 [1], the analysis of large datasets containing biological information has risen in an unprecedented degree. One field that has had a significant development in the last decade is that of proteomics [2]. The function and molecular properties of individual proteins have been studied extensively and information collected was deposited in databases like UniProt [3]. But proteins scarcely ever act individually [4]. Large molecular complexes, formed by interacting proteins perform numerous biological processes vital to the cell's lifecycle. Protein-protein interactions (PPIs) are an integral part of virtually every process that takes place in the human cell [5]. These interactions can be permanent or transient, between homo-oligomers or hetero-oligomers and obligate or non-obligate [6] and can be detected by experimental procedures [7]. Many high-throughput experimental methods are used for the detection of interactions, such as yeast two-hybrid (Y2H) [8], affinity purification mass spectrometry (AP-MS) [9] and protein chip technology [10]. Both low and high-throughput interaction data are deposited in public databases [11]. A compilation of these databases can be found on Pathguide - a meta-database of more than 190 biological pathway and network databases [12]. From all these repositories, the interaction databases most commonly used by the scientific community are DIP [13], MINT [14] and IntAct [15, 16], the main co-founders of IMEx [17], the International Molecular Exchange Consortium. IMEx provides an expertly curated, non-redundant set of molecular interactions from a network of 10 collaborating major public interaction databases. IMEx together with HUPO-PSI (HUman Proteome Organization - Proteomics Standard Initiative) [18] have defined the MIMIx [19] (Minimal Information about a Molecular Interaction) standard, which improves the quality of data and the curation of molecular interactions. Protein-protein interaction networks (PPINs) may help us to have an insight of the cell's functions. In PPINs, proteins are represented by nodes and their interactions by edges in a graphical view. There have been studies to explore human large-scale protein-protein interaction networks [20]. The first efforts to create a systematic map of protein-protein interactions was through large-scale Y2H experiments [21] and recently, efforts have been made to map interactions through next generation sequencing techniques [22]. These and other efforts are combined in the Human Interactome Database [21-23]. Protein Interaction Networks can be of significant value in the analysis of a protein dataset as they may provide a complementary view of the biological pathways in which the proteins participate in and reveal aspects of their functions[24, 25]. However, missing and misleading information through false negatives and false positives that are integrated via various experimental approaches of protein interaction identification can cause problems during the analysis of the results [24].

Membrane proteins are of central importance to the cell as they take part in: ion, metabolite and macromolecule transport across membranes; signal transduction; cell adhesion; cell-cell communication; protein anchoring to specific locations in the cell; regulation of intracellular vesicular transport; control of membrane lipid composition and the organization and maintenance of organelle and cell shape [26]. Membrane proteins can be distinguished based on their association with the membrane: transmembrane proteins span a biological membrane one or more times, lipid-anchored proteins interact covalently with a fatty acid, which anchors to the membrane and peripheral membrane proteins associate with integral membrane proteins or/and the lipid bilayer reversibly. Peripheral membrane proteins are indispensable for the cell's proper function as they have numerous functions; from enzymes and electron carriers to polypeptide ligands (like hormones, toxins and inhibitors) and structural domains [27]. In addition, they are as essential to membrane structure as transmembrane proteins, as their arrangement may affect the membrane conformation, stability, biological activity, folding and the binding of other biomolecules [27]. Peripheral proteins have been shown to have membrane binding domains, a fact consistent with their role in signal transduction and membrane trafficking [28]. These domains are significant for the recognition of lipids and thus allow these proteins to interact with the membrane. Peripheral proteins have also been proposed as potential drug targets mainly due to their interactions with membrane lipids [29]. Moreover, specialized membrane microdomains enriched in sphingolipids and cholesterol, termed lipid rafts, compartmentalize cellular processes and can sometimes be stabilized to form larger platforms through protein-protein interactions [30]. The presence of peripheral membrane proteins in lipid rafts can be of great importance, since proteins in these components take part in endocytosis, transcytosis, signal transduction and receptor recycling among other important cellular processes [31].

      Having such a crucial role, membrane proteins are not studied only as distinct units but as complexes too. Studies for the construction of membrane protein PPINs have been conducted in the past few years [32-34], however, these studies are not focused on human cell membranes but on membranes of other organisms, prokaryotes [32] and eukaryotes [33] as well. The aim of our study was to identify and analyze the molecular interactions of peripheral membrane proteins in order to obtain insights about their role across the human plasma membrane.

## 2. METHODS

### 2.1. Dataset of peripheral membrane proteins

To collect the dataset of human peripheral membrane proteins a search was conducted in UniProtKB/Swiss-Prot [3]. Proteins that in their subcellular location field were experimentally annotated as "peripheral membrane proteins" were collected. Proteins with probable, potential or by similarity confidence regarding that field were excluded from the set. These proteins were grouped based on their presence in various organisms and organelles and a dataset of peripheral membrane proteins from *Homo sapiens* that interact with the plasma membrane was constructed. In addition, an extensive literature search was performed in order to identify additional proteins belonging to this category. The final dataset created includes a total of 277 human peripheral membrane proteins of the plasma membrane (Table S1).

### 2.2. Collection of Protein-Protein Interaction data

Using the Accession Numbers (ACs) collected from UniProtKB/Swiss-Prot for the dataset of 277 peripheral membrane proteins, a query was submitted in IMEx [17]. This query provided a set of interactions for a subset of 249 proteins and only protein interactions between human proteins were used, excluding interaction data with viral proteins. This resulted in a final dataset of 238 peripheral membrane proteins with molecular interactions. The file containing the interactions was formatted in MITAB 2.5 format [35], which describes only binary interactions for one pair of interactors in each row (Table S2). In addition to the network generated using IMEx, a network using interactions deposited in IntAct [15, 16] was created to exclude possible misleading results by low quality interactions. For this reason, all interaction data originated from spoke-expanded co-complexes and with an IntAct MI-score lower than 0.6 were filtered out (Table S3).

## 2.3. Visualization and analysis of the network

For the visualization of the network, both Cytoscape 2.8.3 [36] and Cytoscape 3.0.0 were used, since Cytoscape 2.8 offers a plethora of plugins not yet available for Cytoscape 3.0, whereas the latter provides for several novel and improved options. The dataset of interactions described above was visualized based on the different properties of its proteins, such as their subcellular location.

To perform a graph theory based analysis, the plugin NetworkAnalyzer [37] for Cytoscape was utilized. Using NetworkAnalyzer, a comprehensive set of topological parameters was computed and analyzed. These parameters include the number of nodes, edges, and connected components; the network diameter (the largest distance between two nodes), radius (the minimum among the non-zero eccentricities of the nodes in the network), density (the density of the network's population with edges), centralization (the measurement that shows whether the nodes of the network have on average the same connectivity or a star-like topology), heterogeneity (shows the tendency of a network to contain hubs), and clustering coefficient (a measurement of a graph's tendency to be divided into clusters); the characteristic path length (the average number of connections between nodes, which must be crossed in the shortest path between any two nodes); and the distributions of node degrees, neighborhood connectivities, average clustering coefficients, and shortest path lengths. To investigate the contribution of certain nodes to the network stability attacks were conducted by removing the nodes in descending order of degree and calculating the characteristic path length (CPL) of the network in each case. CPL is commonly used in order to measure the network stability [38-40].

From the interaction data all the 2374 interactors were isolated, and for each protein in that set the subcellular location was examined. All proteins were categorized based on their subcellular location in the following categories: Cytoplasm (396), Endomembrane system (182), Lipid-anchor (25), Membrane (239), Mitochondrion (143), Nucleus (851), Peripheral (242), and Secreted (228) including (39) extracellular Peripheral proteins. For proteins that could not be categorized 7 prediction algorithms (pTarget [41], TargetP [42], WOLF-PSORT [43], BaCelLo [44], PredSL [45], LocTree2 [46] and GOASVM [47] ) were used, in order to include them in one of the categories described above, if three or more algorithms agreed in their prediction. If the subcellular location still could not be predicted that field was characterized as unknown (Table S4). Moreover, using data described in [48], the presence of the network's proteins in lipid rafts was examined (Table S5).

Interactions with drugs for the peripheral membrane proteins of the network were also collected, conducting a search in DrugBank [49, 50]. The 121 drugs collected were categorized in 31 categories according to the field "indications" as described in the DrugCard, thus creating the layer of diseases. The drugs were also classified in 28 groups based on the drug category in which they belong (e.g. anticoagulants). The network created by these data was visualized using Arena3D [51, 52]. Arena 3D uses staggered layers in 3D space, allowing the user to group related data into separate layers; in this case, the proteins, the drugs and the indications/diseases. All data from the drug association analysis are available in Tables S6 and S7.

## 2.4. Clustering and Functional Analysis of the network

ClusterMaker is a Cytoscape plugin that implements several clustering algorithms and provides network views of the results [53]. From the algorithms provided, the MCL (Markov Clustering) algorithm was chosen, in search of clusters that represent macromolecular complexes in the cell [54]. The MCL algorithm appears to be superior in comparison with other clustering methods in detecting clusters in sets of high-throughput interaction data [55, 56]. The clusters detected were compared with known complexes found in the Mammalian Protein-Protein Interaction Database (MIPS) [57] using the Comprehensive Resource of Mammalian protein complexes (CORUM) [58] in order to detect possible novel components of the known complexes. The data for protein complexes deposited in CORUM are manually annotated and data from high-throughput experiments are excluded. For the functional analysis of the network, two (2) online tools that perform GO term enrichment analysis (Gorilla [59]) and GO slim classification (WebGestalt [60, 61]) were used. The analysis was performed for two (2) different datasets: that of the 238 peripheral membrane proteins and, also, the complete set of the network's proteins, 2374 proteins totally, to study the molecular functions and biological processes in which they participate, as well as the cellular components in which they are located. A set of all the network's proteins from which the peripheral membrane proteins had been removed was used as a background set for the functional enrichment analysis of the 238 peripheral membrane proteins in order to examine their function in the human plasma membrane peripherome. All data derived from the clustering and the functional analyses are available in Tables S8, S9 and S10.

# 3. RESULTS AND DISCUSSION

## 3.1 Collection of the Protein-Protein Interaction data

As mentioned above from all the peripheral membrane proteins that were collected, 249 were shown to have interactions in IMEx. These proteins had 4336 interactions isolated from all the databases participating in the consortium (DIP, I2D-IMEx, InnateDB-IMEx, IntAct, MatrixDB, MINT, MolCon, UniProt) except MPIDB, which contains data for microbial proteins and MBInfo, which is a specialized database focusing on mechanobiological interactions. After the removal of non-human proteins and their interactions, 2374 proteins, including 238 peripheral membrane proteins, were shown to have 3445 interactions. We resubmitted a query in IMEx for all 2374 proteins and resulted in a final dataset of 16961 interactions between the 2374 proteins of our network. The 2374 proteins were the nodes and their interactions were the edges that created the network, which was later used in the analysis. The additional network created using data gathered from IntAct consisted of high quality interactions only having 691 nodes and 855 interactions between them.

## 3.2. Analysis of the network

### 3.2.1. Analysis of network structure based on graph theory

For a deeper understanding of the network's functions, its topological parameters were calculated. Topological parameters can be calculated both for directed and undirected networks. In our case, the parameters were calculated for an undirected network (see Methods 2.3). The simple parameters that give certain information for the network are: the clustering coefficient (0.121), the characteristic path length (3.260), the average number of neighbors (13.177), the network density (0.006) and the network heterogeneity (1.765). Moreover, the complex parameters were calculated in order to obtain a better view of the network. Such parameters are the node degree distribution and the average clustering coefficient (Fig.1). By examining the network's properties one-by-one, interesting information can be obtained. Starting with the network density, it was observed that it has a low value (<0.1) [62]. This is a characteristic often present in many biological networks and it has been argued that biological networks are generally sparsely connected, as this confers an evolutionary advantage for preserving robustness [63]. Another important measure is the clustering coefficient. In random networks, the clustering coefficient is approximately 1/N, where N is the number of the nodes of the network [64]. Biological networks have significantly higher clustering coefficients compared to random ones -as is the case here-, which shows that cellular processes are executed by subsets of molecules forming functional modules [63], as seen during the MCL clustering process.

The understanding of the topology of a network can give insights relevant to its biological significance. The basic parameter that reveals the topology of a network is the node degree distribution (Fig. 2A). In our case the distribution is of the form:

$$P(k) = 1015.8 k^{-1.383} \qquad (3.1)$$

In scale-free networks the probability P(k) that a vertex in the network interacts with k other vertices decays as a power law, following P(k) ~$k^{-\gamma}$ - as is the case here - where γ is the degree exponent [65]. The value of γ determines important properties of the network. In cases where the value of γ<2, the role of the hubs in the network becomes more important, than in most cases where 2<γ<3 [66]. Biological networks are robust against random node failures, but disruption of hubs (proteins with a large number of links) often leads to system failure [67]. Scale free networks have average path lengths significantly smaller than that of random networks. To compute the average path length for a random graph we used the formula introduced by Fronczak et al. [68]

$$l_{random} = \frac{\ln N - \gamma}{\ln\langle k \rangle} + \frac{1}{2} \qquad (3.2)$$

where γ=0.5772 is Euler's constant. For this case, the average path length is 7.55, which is larger than 3.26 – the characteristic path length of the network. From these data we can observe that the network has a scale-free topology, where the hubs hold the network together [69]. Proteins with high node degrees (>30) are considered hubs in the network. These proteins have various functions: from receptors for GABA (UniProtKB AC:Q9H0R8) and estrogens (UniProtKB AC:P03372) to structural proteins like fibronectins (UniProtKB AC:P02751) and microtubule associated proteins (UniProtKB AC:Q9H492), all important for the sustention of the cell's homeostasis and precise function. For instance, knockout of the von Hippel-Lindau tumor suppressor, a peripheral membrane protein (UniProtKB AC:P40337), which is also a hub in our network, causes prenatal lethality due to abnormalities in the morphology of various organs during the development and organogenesis of the embryo in mice [70-72].

As stated above, for random networks the clustering coefficient is C ~ 1/N, and in this case C ~ $4.21 \cdot 10^{-4}$ which is very small compared to the clustering coefficient of the network (0.121). This combined with the fact that the average path length of our network is small compared to that of a random network, indicates a small-world network [73]. Small-world networks can efficiently transmit information between distant nodes (small path length), while simultaneously process local information efficiently (high clustering coefficient) [74]. For example, in our network, a peripheral protein with a high clustering coefficient and a small average path length is the subunit sigma of AP-2 complex (AP2S1) (Fig.S1). This protein is part of the adaptor protein complex 2, which functions as a protein transporter in different membrane traffic pathways via transport vesicles and is involved in clathrin-mediated endocytosis. This protein interacts in our network with three (3) other proteins -Epidermal growth factor receptor substrate 15 (EPS15), Growth factor receptor-bound protein 2 (GRB2) and Vascular cell adhesion protein 1 (VCAM1)-, which take part in signal transduction and cell-cell recognition. These proteins in turn interact with 510 other proteins in the network. This signifies AP2S1 as a "bottleneck" and a very important protein for the network's normal function and subsequently the cell's vitality. Its removal would destroy many links between the 510 proteins that are now connected. Interestingly, knockouts of GRB2 and VCAM1 lead to pre-natal lethality in mice [75, 76]. Using data collected from the Mouse Genome Database (MGD) [77] we were able to characterize the hubs and bottlenecks of the network as proteins encoded by essential (knockout of these genes in mice produce lethal phenotypes) and non-essential genes for the organism's development (Table S11). We observe that 52% of hubs in the network of the plasma membrane peripherome, and 25% of bottlenecks are essential proteins as knockouts of their protein-coding genes results in abnormal survival (lethality) in mice. More details about the top 10 hubs and bottlenecks of this network are presented in Table S12.

To examine whether a protein shares interaction partners with other nodes in the network the topological coefficient has to be measured [78]. The topological coefficient decreases with the number of links (Fig. 2C), which is an indicator that hubs in the network are as connected as the rest of the network's proteins, thus suggesting that hubs in the network are not clustered together. It also indicates, in compliance with the clustering coefficient, that the network has a modular organization [25].

The gradual removal of proteins present in lipid-rafts and peripheral membrane proteins of the human plasma membrane in descending order of node degree ('attacks' [40]) caused a more rapid increase in the characteristic path length (CPL) of the network compared to the removal of the same proportion of randomly selected nodes from the whole network (commonly described as 'failure' [40]) (Fig.3). The increase in the network's CPL shows the importance of the removed proteins as mediators of intracellular communications, since the paths connecting the remaining nodes in the network are longer (Table S13). This effect on the CPL for peripheral membrane proteins and lipid raft related proteins in the human plasma membrane peripherome indicates their importance for the stability and proper function of this cellular sub-network.

The network structure and topology is similar for the network created using only high-quality data present in IntAct. P(k) decays as a power law, following P(k) ~$k^{-\gamma}$ ($P(k) = 474.7k^{-2.036}$), the clustering coefficient of the network is larger than that of a random network and the average path length is smaller. The IntAct network has a scale-free topology, small-world properties as well and the complex parameters of this network have similar distributions to those mentioned for the network created using data from IMEx, thus suggesting that the differences between the two networks are not such to suggest that the presence of data from high-throughput experiments have a severe effect on the networks topology and characteristics.

### 3.2.2. Analysis based on the proteins' subcellular location

All the network's proteins were categorized based on their subcellular location (Fig. 4). The main observation was that peripheral proteins interact with proteins that have multiple subcellular locations and are not mainly localized in the plasma membrane as perhaps expected. Peripheral membrane proteins of the plasma membrane constitute a central node connecting the plasma membrane with the entire cell, as they can detach themselves from the membrane plane and interact with proteins with multiple subcellular locations inside the cell.

Through this transient membrane association, they are involved in protein transport and are essential for the regulation of vesicle transport [79]. This is indicative of the central importance of the peripheral membrane proteins, as they constitute a sub-proteome, pivotal for the cell's metabolism and function. Another observation made was the preference of extracellular peripheral membrane proteins to interact with transmembrane proteins, and especially single-pass transmembrane proteins of the plasma membrane. The majority of single-pass transmembrane proteins belongs to type I (proteins spanning the membrane once, with their N-terminal on the extracellular side of the membrane and their signal peptide removed) and are associated with the immune system. This is consistent with the fact that the secreted peripheral proteins are mainly associated with inflammatory processes and so it is logical for these proteins to interact with other proteins of the immune system like interleukins, antigens and cytokines. Approximately 50% of transmembrane proteins of the human plasma membrane peripherome act as receptors and almost a quarter of them present catalytic activity (Table S15). Interestingly, 182 (ca.8%) of proteins in the human plasma membrane peripherome are located in lipid rafts and 45% of them have catalytic activity while 29% act as receptors. The distribution of the proteins amongst the various subcellular locations is not affected by the removal of "low quality" interactions - as shown through the analysis of their distribution in the network composed from data collected from IntAct (Table S14).

### 3.2.3. Analysis of peripheral membrane proteins' interactions with drugs

From the data collected from DrugBank [49, 50], a correlation between drugs and diseases was made based on the field "Indications". Peripheral membrane proteins and drugs interacting with them are associated mainly with cardiovascular and blood associated diseases, and cancer. There are also a few proteins associated with asthma, arthritis and skeletal disease. It was observed that ca. 10% of the peripheral membrane proteins in the plasma membrane are targets for ca. 10% of approved drugs. But if the peripheral membrane proteins are examined based on their subcellular location the case is different, since 40% of extracellular peripheral membrane proteins have interactions with at least one drug. There is also a correlation of "drug category" (as described in the Drugbank card) and the proteins' subcellular locations: the majority of the drugs belong to a category in which all of them interact with proteins that are either in the cytoplasmic (e.g. drugs that belong to the category of contraceptives) or the extracellular (e.g. drugs that belong to the category of anticoagulants) side of the plasma membrane. Interactors (Fig. S2) are proteins in the network, which interact with the peripheral proteins of the plasma membrane that are associated with drugs (as described above). These interactors are categorized - based on their association with the membrane plane - to six distinct categories: multi-pass transmembrane proteins, single-pass transmembrane proteins, peripheral membrane proteins, secreted proteins, intracellular proteins and other. It was observed that extracellular peripheral membrane proteins interact mostly with single-pass transmembrane proteins and secreted proteins. This was also observed for all the extracellular peripheral proteins mentioned in the previous section. A certain example for Alzheimer's disease is given in Fig. S2, where the drugs associated with this disease, the proteins connected with these drugs and their interactors are selected.

### 3.2.4. Functional Analysis of the Network

From the GO term slim classification it is apparent that the 2374 network's proteins take part in metabolic processes (especially protein modification), biological regulation (especially intracellular protein kinase cascade), apoptotic processes and intracellular transport. The cellular components in which most proteins are located in are the cellular membranes and the nucleus, a fact additionally present in their categorization based on subcellular location (Fig. 4). As for the protein molecular functions all the network's proteins take mainly part in protein (especially ubiquitin protein ligase and nuclear hormone binding) and ion (ATP) binding and present catalytic activity (protein kinase), but peripheral membrane proteins take part in lipid binding to a much larger degree than the rest of the proteins in the network. This is relevant with the fact that the majority of peripheral proteins that interact directly with the membrane have lipid binding domains [80].

More detailed results were gathered through the GO term enrichment analysis conducted using GOrilla [59]. In the results of the enrichment analysis every biological process, molecular function and cellular component is associated with certain proteins in the network and a P-value is given to each association according to its significance (lower values correspond to larger significance). Processes with fairly low P-values ($10^{-10}$ – $10^{-6}$) for all the network's proteins present in the results of the analysis were ATP synthesis coupled proton transport, cholesterol efflux and TAP-independent antigen processing and presentation of exogenous peptide antigen via MHC class I. However, only a small number of proteins is associated with these processes (up to 1%), thus implying that they could belong to a certain cluster of the network and that they are not representative of the process that the whole network is conducting. Applying the same logic for molecular function and cellular components we were able to isolate such examples like G-protein coupled amine receptor activity, MHC class I receptor activity and translation factor activity for molecular function and triglyceride-rich lipoprotein particle, clathrin adaptor complex and AP-type membrane coat adaptor complex for cellular components. For the peripheral proteins in particular there was an evident association with the plasma membrane and the cytoskeleton. As noticed during the slim classification process as well, peripheral proteins participate in lipid binding and especially phospholipid binding and regarding biological processes, those showing an overrepresentation are the regulation of signaling, cell communication and phosphate metabolic processes.

### 3.2.5. Network Clustering

For the detection of macromolecular complexes in the network, the MCL clustering algorithm was used (see Methods). The inflation parameter was set to 1.8. This value has been shown to give the best results regarding the identification of functional modules in PPINs [55]. The algorithm detected 160 complexes in total. For these complexes we performed functional analysis using the WebGestalt toolkit, in order to detect biological similarities between the proteins in each complex. We selected the human genome as a reference gene set in order to perform enrichment analysis to obtain information from all the databases in WebGestalt (see section 2.4 above). We evaluated the data obtained from these analyses for all the protein clusters at hand, and assigned a specific biological activity or disease to those complexes that a significant correlation with specific terms could be made. This allowed us to retrieve 45 macromolecular complexes with a certain function. Some of these complexes are depicted in Fig. 5. The complexes in the network take part in a long range of processes inside the cell, from cell division to the regulation of matrix adhesion and antigen processing and presentation, thus revealing the central role of peripheral proteins in multiple cell functions. In Fig. 5B some of the protein complexes that have a relation with diseases are shown. The diseases with which the clusters' proteins associate are mainly infectious diseases (e.g. AIDS), cardiovascular and blood diseases and disorders (e.g. hypertrophic cardiomyopathy and polycythemia vera) and cancers (e.g. acute myeloid leukemia). The peripheral membrane proteins in these complexes are of central importance and further drug research should focus on them as they appear to be central nodes connecting proteins related to the same disease.

All 45 complexes indentified were compared with known complexes in MIPS [57]. 25% of these complexes were partially correlated with complexes deposited in MIPS. Two characteristic examples are shown in Fig.S3. The first is the endosomal sorting required for transport complex ESCRT-III, which is required for intracellular transport [81]. Seven proteins of this cluster are known components of this complex. There is a potential novel member of the ESCRT-III complex present in this cluster. It is a Multivesicular Body Protein (MBP) involved in BD formation, a function conducted by ESCRT-III complex consisting this protein a probable core component of ESCRT-III [82]. The other proteins in this complex are associated with other ESCRT complexes [83] or cell trafficking in general [84] and could possibly be components of this complex as well since their function has not been studied extensively. The second is the mTORC2 complex, a protein complex regulating the cytoskeleton. Two of the proteins not currently in the complexes in MIPS are described as potential members of the mTORC2 complex [85, 86] (Fig.S3). In addition, all complexes were examined for the existence of peripheral proteins and proteins located in lipid rafts (Table S5).

## 4. CONCLUSIONS

Almost 50% of human proteins are intrinsic or peripheral to cellular membranes [80] and are one of the most important components of the human cell. In this work we studied the interactions of peripheral proteins of the human plasma membrane and analyzed various properties of the interaction network in order to obtain a better understanding of the characteristics of the human plasma membrane peripherome. We observed that the network possesses the characteristics of biological networks, thus having a scale-free topology and small world properties. Further analysis, indicated proteins which are essential for maintaining the connectivity and stability of the network. Removal of these proteins leads to deleterious effects on cells and lethal phenotypes in mammals. Peripheral membrane proteins and proteins of the network located in lipid rafts are important for the stability of the human plasma membrane peripherome and have a greater contribution to the stability of the whole network than other proteins since their removal leads to network destruction. Peripheral membrane proteins are targets for ca.10% of approved drugs and are associated with multiple diseases. There are also other proteins in the network, belonging to certain complexes, associated with diseases that are commonly studied and have high prevalence in the human population. The plasma membrane peripherome participates in cell trafficking, signal transduction and apoptotic processes and is consisted of proteins present in almost every subcellular compartment. Certain examples presented here underline the importance of these proteins in the formation and functionality of the biological system they forge since their removal can even lead to organismal lethality (e.g. von Hippel-Lindau tumor suppressor [70-72]. The presence of lower quality interaction data can cause problems during the analysis of the network. For this reason we created and analyzed an additional network consisting of high quality data only and observed that the basic characteristics of the network remain the same. Protein-protein interaction data for membrane proteins are under-represented in public databases but their study can reveal important features of membrane proteins and guide future experiments. In conclusion, this study reveals certain new properties and features of the peripheral membrane proteins and the network created from them and their interactors; hence their study uncovers their pivotal role in the cell's functionality and vitality. The study of the human plasma membrane peripherome reveals potential candidates (e.g. hubs and bottlenecks) that can be used for further experimental studies. Moreover, the study of various human cell interactomes – such as the one studied here – can be of great importance for the mapping and the identification of the complete human cell interactome and as Dennis Bray stated "*we have a new continent to explore and will need maps at every scale to find our way*" [87].

# Acknowledgements

*Funding*: The present work was funded by the SYNERGASIA 2009 PROGRAMME, co-funded by the European Regional Development Fund and National resources (Project Code 09SYN-13-999), General Secretariat for Research and Technology of the Greek Ministry of Education and Religious Affairs, Culture and Sports.

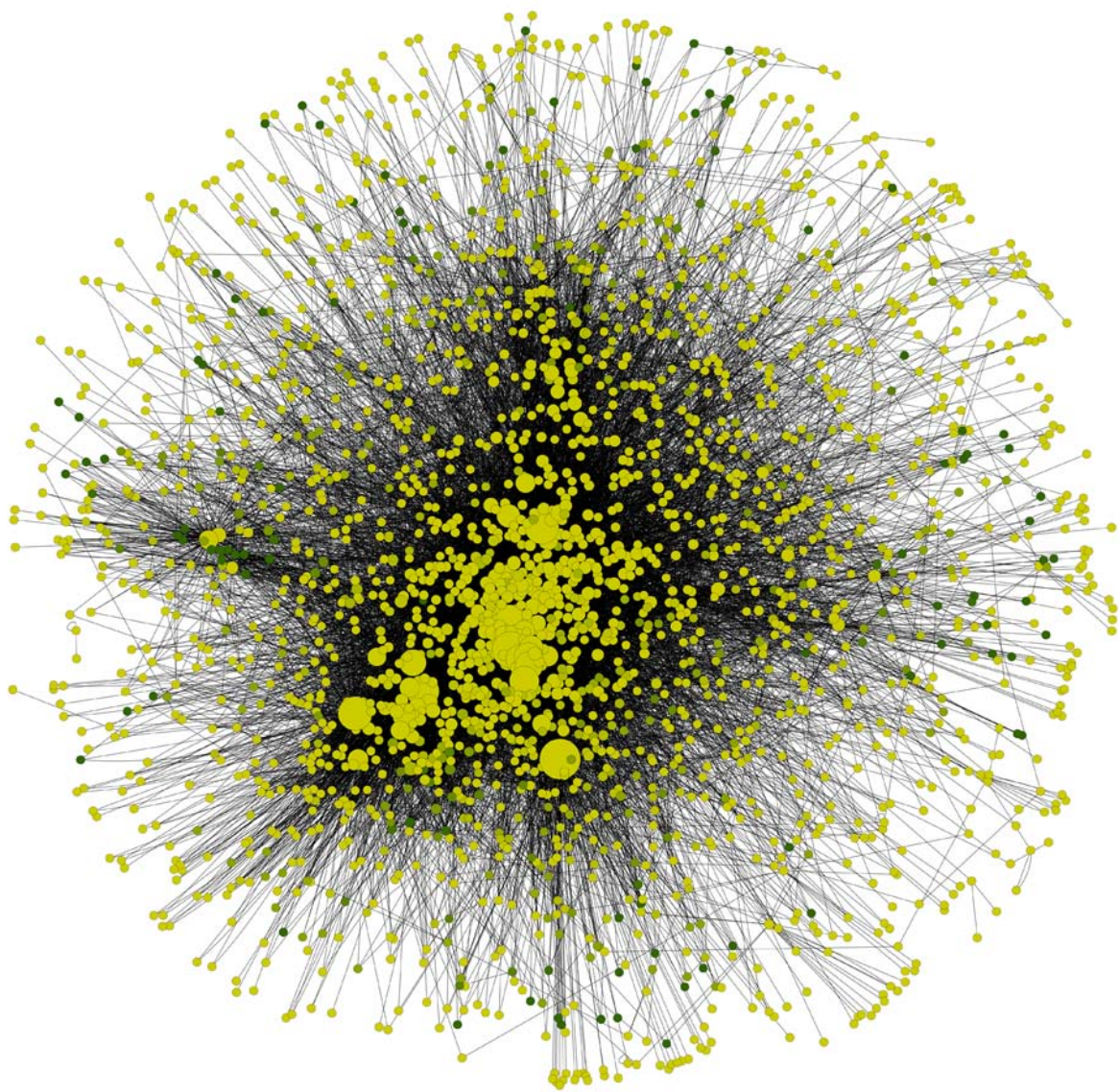

**Figure 1.** Visualization of graph theory parameters in the network of the human plasma membrane peripheral proteins with 2374 nodes and 16961 edges. The colour gradient is visualized based on the clustering coefficient of each node. In darker colours are the nodes with the higher clustering coefficients grading to lighter colours for nodes with lower clustering coefficients. A size gradient is used to map the node degree on the network's proteins. Larger nodes are indicative of nodes with higher degrees and smaller nodes of nodes with lower degrees.

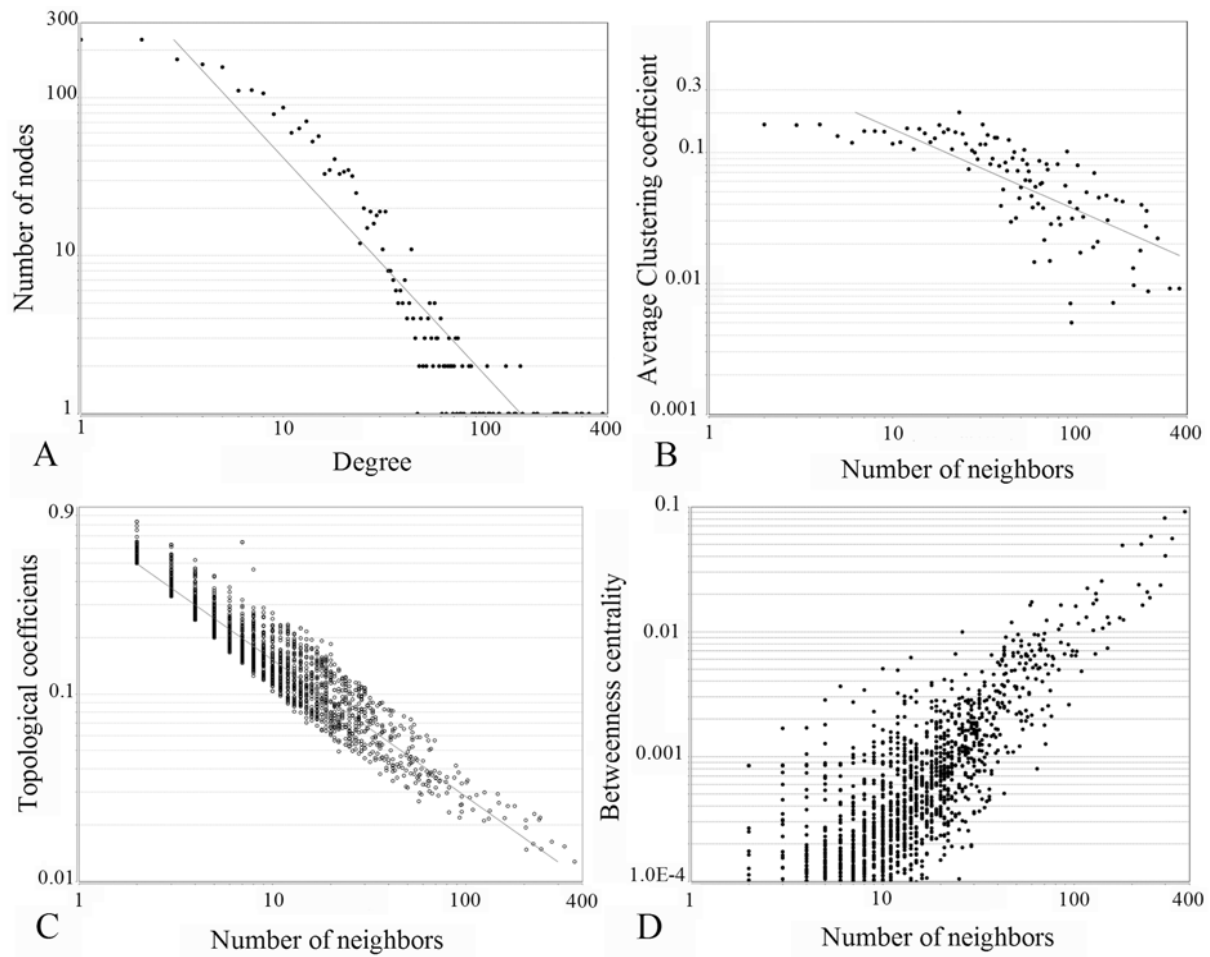

**Figure 2.** Charts for four complex topological parameters of the network of the peripheral proteins of human plasma membrane. **A.** Node degree distribution which decays as a power law and accentuates the scale-free properties of the network. **B.** Average clustering coefficient distribution. **C.** Topological coefficient distribution. Both the average clustering coefficient and topological coefficient distributions indicate that the network has a modular organization. **D.** Betweenness centrality distribution. This distribution shows that proteins with high betweenness centralities, which act as hubs are few, compared to the rest of the proteins in the network. All the distributions shown above are in log-log plots.

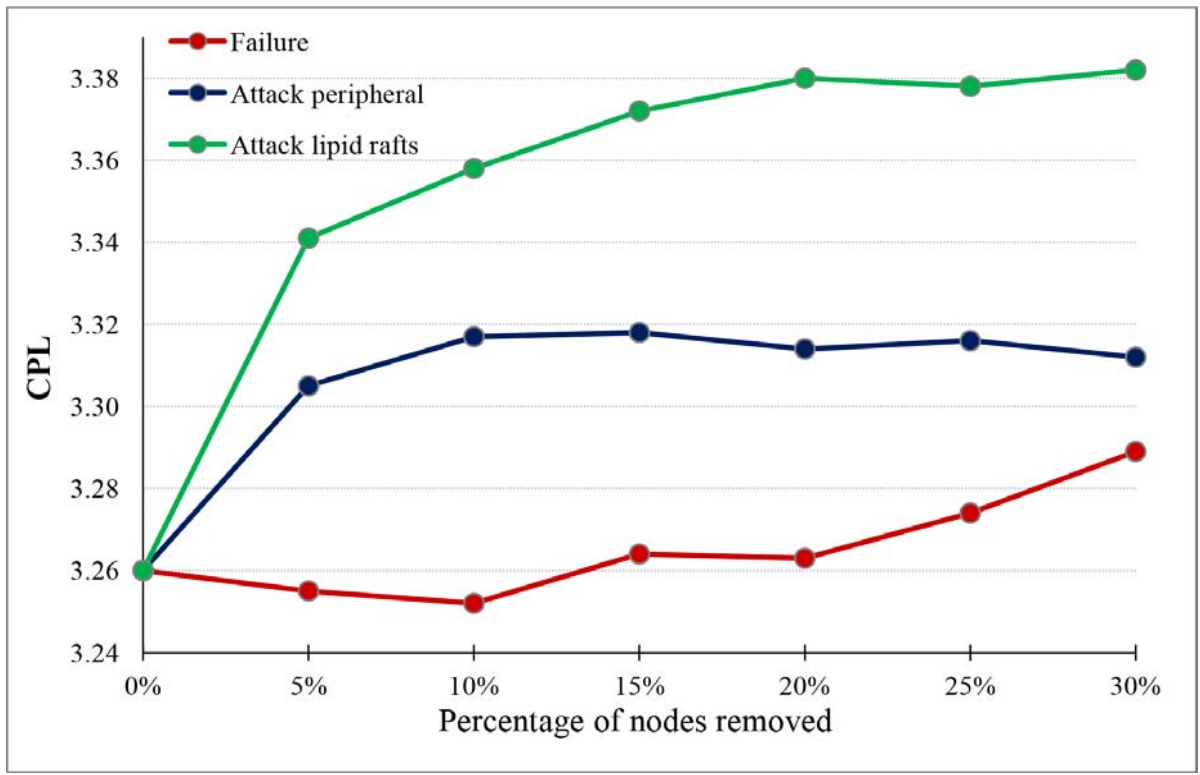

**Figure 3**. The effect of the gradual removal of selected (attacks) and random (failure) nodes to the characteristic path length (CPL) of the network. The removal (attack) of peripheral proteins of the human plasma membrane and proteins located in lipid rafts based on their node degree causes a more rapid increase of the characteristic path length of the network compared to the removal of random nodes (failure).

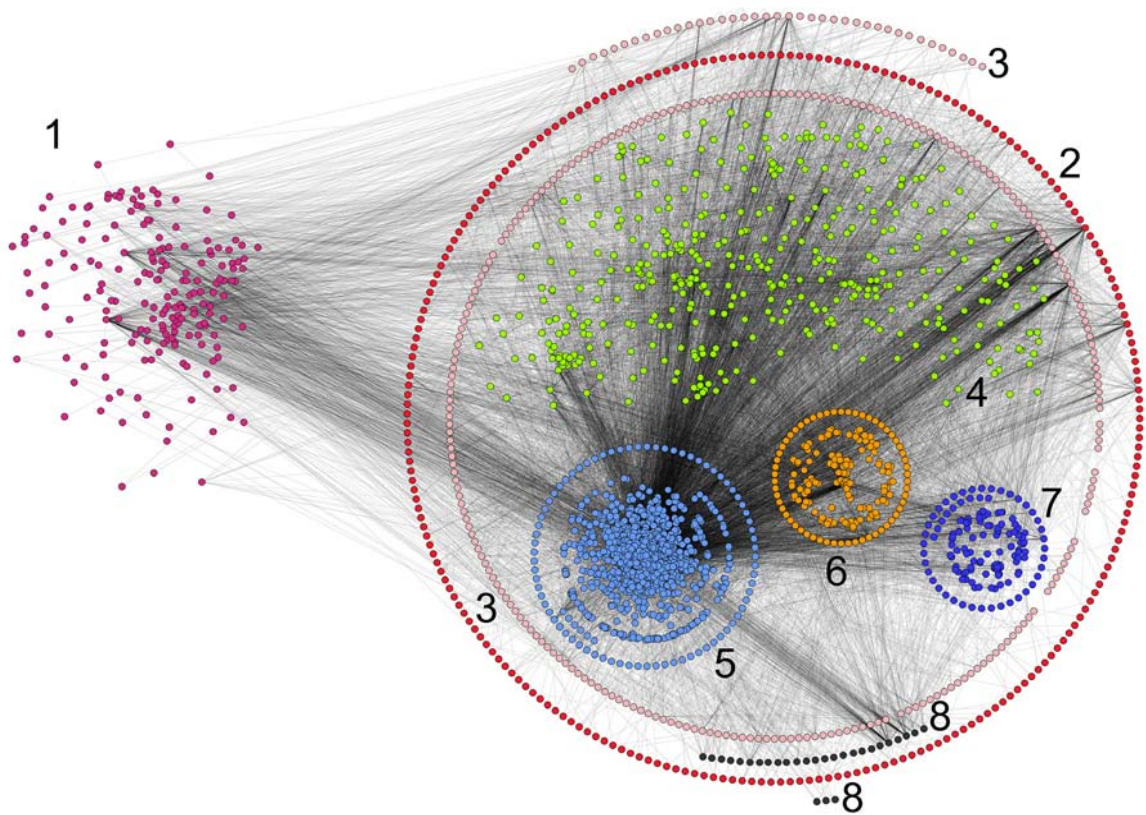

**Figure 4.** Proteins are grouped based on their subcellular location in the network of the human plasma membrane peripheral proteins. *1:* secreted proteins, *2*: membrane proteins, *3:* peripheral proteins, *4:* cytoplasmic proteins, *5*: nuclear proteins, *6*: endomembrane system proteins, *7*: mitochondrial proteins, *8*: lipid-anchored proteins. The membrane proteins of organelles and the endomembrane system are forming circles around the other proteins of the organelles. Proteins with unknown subcellular location are hidden in this view of the network.

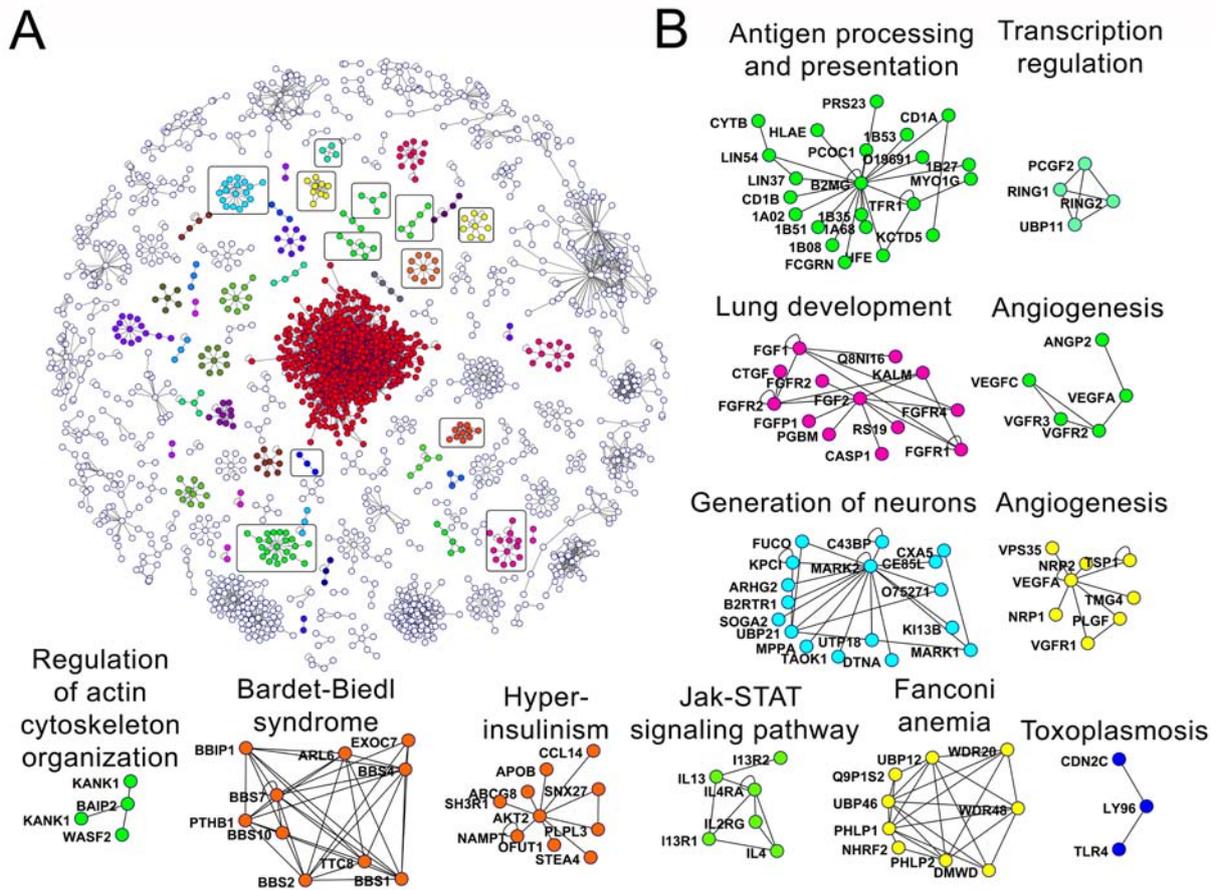

**Figure 5. A.** All the 160 complexes produced from the MCL clustering process are shown. In white are complexes that no biological significance could be detected and in the remaining colours are complexes that are associated with a certain biological function or disease. **B.** In this figure the complexes that are circled in A are shown. These complexes are some characteristic examples of complexes with a certain biological function or disease. The gene name that maps to each protein is shown for all the nodes in each cluster. Every complex has a label consistent with its association.

# The Human Plasma Membrane Peripherome: Visualization and Analysis of Interactions

Katerina C. Nastou, Georgios N. Tsaousis, Kimon E. Kremizas, Zoi I. Litou and Stavros J. Hamodrakas[*]

Department of Cell Biology and Biophysics, Faculty of Biology, University of Athens, Panepistimiopolis, Athens 15701, Greece

# Supporting Information (1)

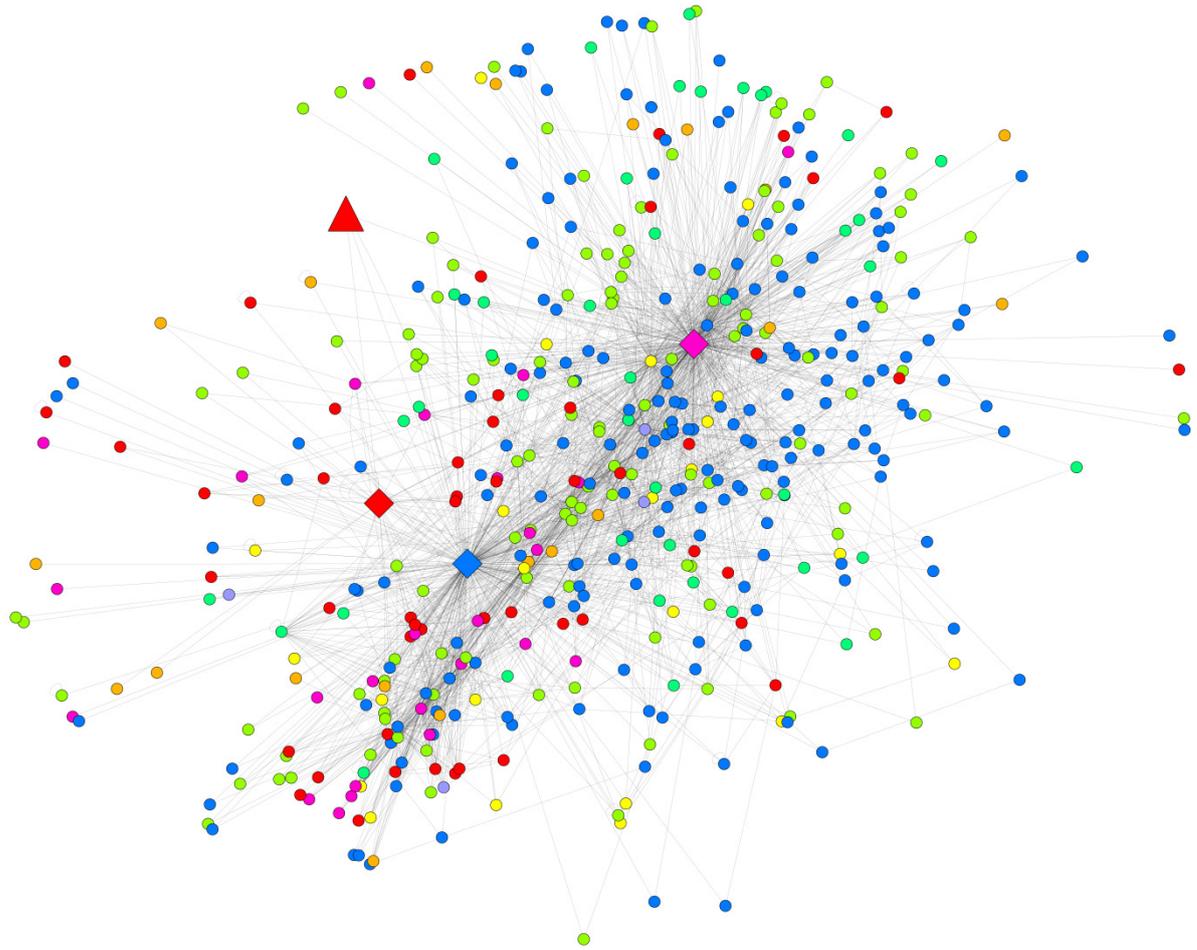

**Figure S1.** The sub-network of the sigma 1 subunit of the adaptor related protein AP2S1 (red triangle), a bottleneck, connected with 3 other proteins (magenta, red and blue diamonds) which are further connected with 510 other proteins in total.

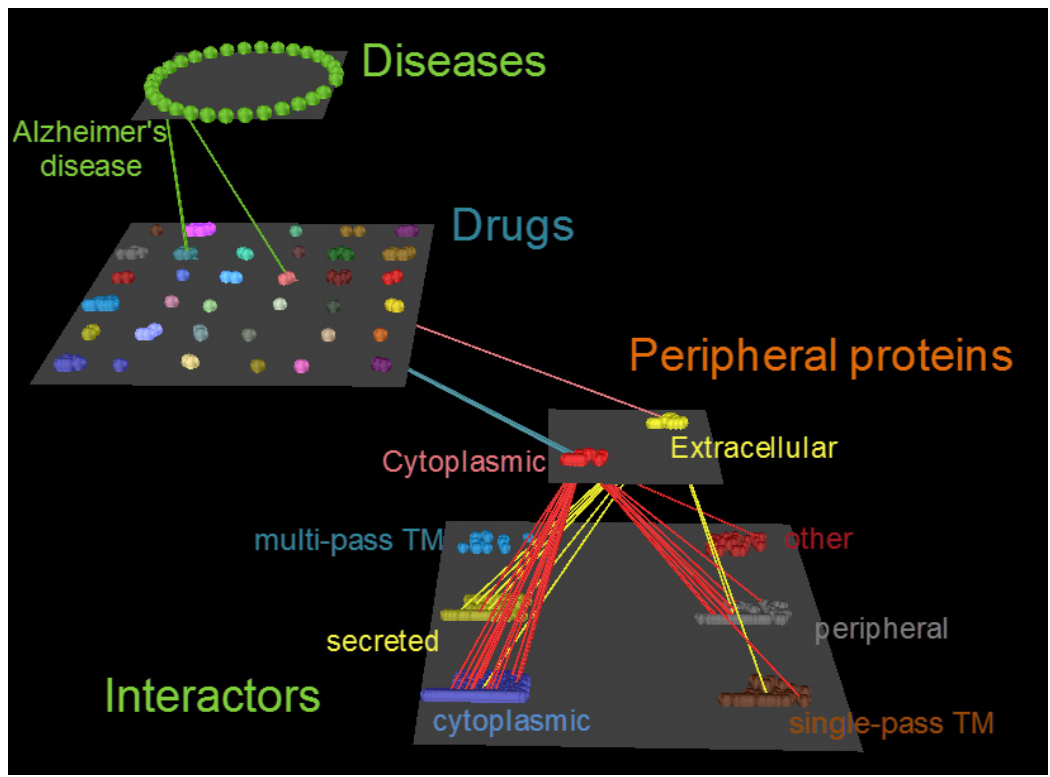

**Figure S2.** A four layered network created with Arena3D is shown. The first layer (Diseases) consists of 31 diseases which are connected with 125 approved drugs grouped in 28 different drug categories as collected from DrugBank, that create the second layer of the network (Drugs). These drugs interact with 31 peripheral membrane proteins, which constitute the third layer of the network and are grouped based on their subcellular location. In the fourth layer (Interactors) the proteins that interact with the peripheral proteins are depicted, creating the layer of interactors. These protein interactors are grouped based on their association with the membrane plane in six categories (multi-pass transmembrane proteins, single-pass transmembrane proteins, peripheral membrane proteins, secreted proteins, intracellular proteins and other). An example of the connections between Alzheimer's disease and the drugs associated with it (DrugBank IDs: DB00163, DB00144, DB02381), the peripheral proteins of the human plasma membrane interacting with these drugs (Q16760, P17252, Q8N3E9, O95810, P15692) and the interactors of these proteins.

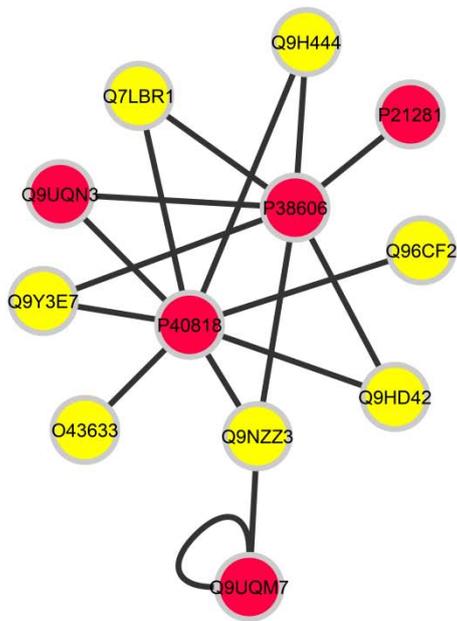 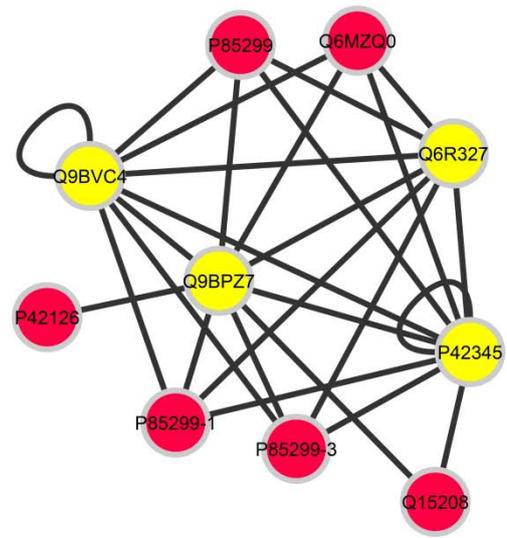

ESCRT-III complex                     mTORC2 complex

**Figure S3.** Two complexes with novel components produced from the MCL clustering process are shown. The first complex is the endosomal sorting required for transport complex III (ESCRT-III complex). The second complex is the mammalian target of rapamycin complex 2 (mTORC2) which functions as an important regulator of the cytoskeleton. Proteins that represent previously reported components of the complex are shown in yellow, while novel subunits are shown in pink.